# On the importance of functions in data modeling

Alexandr Savinov

http://conceptoriented.org

31.12.2020

**Abstract.** In this paper we argue that representing entity properties by tuple attributes, as evangelized in most set-oriented data models, is a controversial method conflicting with the principle of tuple immutability. As a principled solution to this problem of tuple immutability on one hand and the need to modify tuple attributes on the other hand, we propose to use mathematical functions for representing entity properties. In this approach, immutable tuples are intended for representing the existence of entities while mutable functions (mappings between sets) are used for representing entity properties. In this model, called the concept-oriented model (COM), functions are made first-class elements along with sets, and both functions and sets are used to represent and process data in a simpler and more natural way in comparison to purely set-oriented models.

## 1 Introduction

The main purpose of a *data model* is to provide a formal definition of what we mean by data at some level of abstraction, how we represent data and how we process data. In other words, if somebody asks what is data then a data model should give clear answers to these questions. Typically, a data model includes a definition of data elements (data representation), their organization by describing how they relate to each other (data structure) as well as how the elements and the relations are interpreted in terms of real-world entities (data semantics).

There exist many generic approaches to data modeling and numerous variations depending on how data elements are defined, how they are structured, how they are processed and how they are semantically interpreted. However, most of them assume that the minimum unit of data is a *value* as it is understood in computer science, that is, something unique, immutable and what can be passed only by copying its whole content (hence the term copying by-value).

Formally, values can be modelled by mathematical *tuples* which are interpreted as combinations of other values. This basic structure of tuple membership probably exists in all data models as well as other branches of computer science. The idea is that given two or more tuples, we can create a new tuple which combines them into one construct. Note again that such combinations are created by copying the member values and a tuple *is* a (structured) value.

Although tuples and tuple membership are at the core of most existing data models, they are not enough for a complete data model. The second basic structure widely used in data modeling is intended to describe collections of values. Such collections are formally represented by mathematical *sets* the members of which are tuples. Data manipulations in this case are reduced to adding tuples to a set or removing them from a set. Note that tuples (data values) are not sets and hence this model does not allow us to represent nested set membership. Theoretically, it is possible to assume that sets may include other sets as its members but for certain reasons this approach did not find wide acceptance.

Having tuples (structured values) and sets (collections of tuples) is already enough for creating a full featured data model. We refer to this large category of models as *set-oriented models* because they rely on manipulating tuples in sets. In particular, the relational model (RM) of data [2] belongs to this category but also less formalized approaches to data representation and processing like no-SQL databases or map-reduce assume that data is represented by collections of elements and processed by producing new collections from existing collections.

One important property of this generic set-oriented setting is the usage of tuple attributes for representing entity properties. The currently dominating assumption is that entity properties are mapped to tuple attributes, that is, tuples have as many attributes as the represented entity has properties. For example, if a product item is characterized by name and price then it is represented by a tuple with at least two attributes.

However, this assumption is quite controversial and essentially contradicts to some basic principles of set-orientation. The main issue with this assumption is that tuples are by definition immutable and hence it is impossible to modify entity properties without a conflict with set theory. The issue could be resolved by adding some stipulations, assumptions or additional mechanisms like primary keys but they will only hide the problem and not eliminate it. Essentially, such additional mechanisms as primary keys are a workaround which makes the model more complicated without the resolution of the problem at basic level.

In this paper, we describe a solution to this problem which is an alternative view on how data can be modelled at fundamental level. More specifically, instead of using tuple attributes, we propose to use mathematical *functions* to represent mutable entity properties. For example, if a product item is characterized by name and price, then we do not define tuple attributes for their representation. Instead, we define two functions: the first function maps product items to names and the second function maps product items to prices. Note that functions are fundamentally different from tuple attributes because they are mutable and support two basic operations: setting (assigning) a property value and getting (reading) a property value.

The goal of the paper is to show that functions are better for modeling (mutable) entity properties than tuple attributes. This solution makes data modeling simpler and more natural due to having much less semantic load on tuples and sets. In general, we assume that entity identifiers are not entity properties and hence they have to be modelled using different constructs: sets and functions, respectively. The usage of functions for data modeling is of course not new [4, 10] but previously they have been used as an additional layer over sets and set operations, or for conceptual modeling. In these cases, the whole model still heavily relied on set theory, particularly, by expressing all data manipulations in terms of set operations. The model we describe, called the concept-oriented model (COM) [5], makes functions equal to sets by moving many mechanisms from sets to functions. In particular, tuples are not used for representing entity properties. Inference in COM relies not only on set operations but also on operations with functions (deriving new functions from existing functions). This changes the way data is processed and this approach was implemented in several systems such as Prosto[1], Lambdo[2], Bistro[3] for general purpose data processing and DataCommandr [6] and ConceptMix [8] for data wrangling.

In Section 2, we introduce some basic notions and describe how data is modeled in set-oriented models. In Section 3, we describe the problem of modeling entity properties using tuple attributes. Section 4 describes how the mechanism of primary keys is used to solve this problem. Section 5 introduces functions for data modeling and demonstrates how they can be used to solve the problem of modeling entity properties. Section 6 describes main properties of the concept-oriented model (COM) which relies on both sets and functions for data modeling. Section 8 provides concluding remarks.

## 2 Representing entity properties by tuple attributes

One of the main assumptions of set-oriented models including the relational model (RM) is that entities of the real world are formally represented by mathematical *tuples*. This means that if we want to represent a new entity then we create a tuple in some set. If we do not need to manage this entity anymore then this tuple is removed from the set. Adding and removing tuples from a set are basic operations of any set-oriented model, that is, all possible data manipulations are reduced to these two operations. Since tuples are (structured) values, this approach means that one entity is represented by one value.

Now let us consider how the *structure* of entities is represented by elements of the model. Here we use a general informal observation that entities have *properties* which allow us to characterize an entity.

---

[1] https://github.com/asavinov/prosto
[2] https://github.com/asavinov/lambdo
[3] https://github.com/asavinov/bistro

The question is then how entity properties are represented in a set-oriented model which provides only tuples in sets for data representation. The dominating pattern here is that entity properties are represented by tuple *attributes*, that is, entity structure is mapped to tuple attribute structure. For each new entity, we need to analyze its internal structure and then create one tuple attribute for each property which is supposed to contain the value characterizing this entity. Tuples and tuple attributes are viewed as a "natural way of representing properties of an entity" [1] also in other models frequently just because there are no other constructs except for sets, tuples and tuple attributes.

The principle of representing entities and their properties by tuples and their attributes, respectively, is at the core of set-oriented data modeling as well as many other approaches. For example, let us assume that a product is characterized by its name and price. In order to represent such an entity, we define two attributes: name attribute and price attribute (Fig. 1). We would need also some kind of identifying property which is also represented as a tuple attribute. In Section 5, we will describe an alternative approach where neither entity properties nor entity identifiers are tuple attributes.

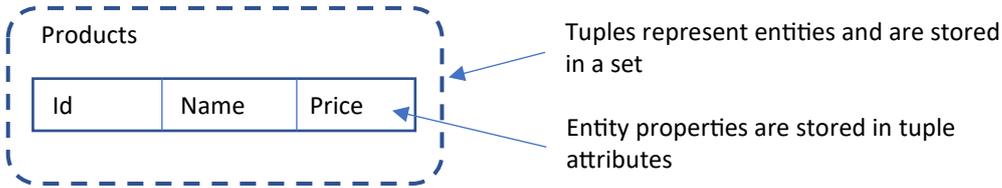

*Fig. 1 Representing entities by tuples and their attributes in a set-oriented model*

## 3  Problem: Changing the unchangeable

In this section we show that the usage of tuples and tuple attributes for representing entities has one serious flaw. Assume that some property of an entity has changed and we want to update its representation to reflect this change. In our case, if properties are represented by tuple attributes, we need to accordingly modify the attribute corresponding to this changed property.

It seems that the problem does not exist because we can simply modify the corresponding attribute in the tuple. In particular, the relational model provides UPDATE operation for that purpose (in addition to INSERT and DELETE set operations) which allows us to change tuple attributes. Note that although RM recognizes and frequently explicitly stipulates that values cannot be updated, the mechanism at logical level works exactly as if attributes were mutable, and the existence of the UPDATE operation manifests such a possibility.

Although practically it is possible to modify tuples to reflect changes in entities, formally it is a workaround or even formal hack. Indeed, if we remember that tuples are *by definition* immutable then we get a conflict with one of the main assumptions. It is simply not allowed to modify tuples in any model which respects the principles of set theory. It is only possible to add a new (unique) tuple and remove an existing tuple – nothing else. Thus, strictly speaking, there is even no question how good the mechanism of using tuple attributes for representing entity properties is – all of them will have immutable entities (if they want to be referred to as set-oriented approaches). Such a model with immutable entities can hardly be useful and hence some solution is needed.

One simple workaround (accepted in RM) consists in removing a tuple and then adding a new modified tuple. For example (Fig. 2), if a product changes its price, then we need to delete the tuple <"My Product", 12.34> and add a new tuple <"My Product", 23.45, >. Note that this approach means removing a whole entity along with its identifier and all properties, and then adding a new entity with the same identifier and modified attributes.

Removing and adding a whole tuple instead of modifying individual attributes is a kind of a hack similar to justifying the use of non-unique tuples in sets. Another workaround is described in the next section.

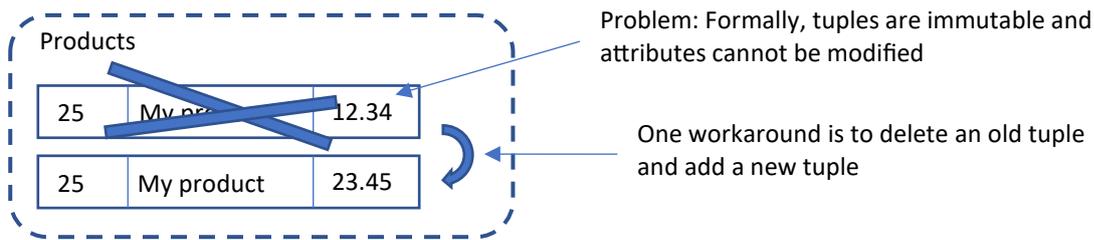

*Fig. 2 Tuples are immutable and one way to modify them is to remove and add a new tuple*

## 4 Workaround: Declaring immutable as mutable

An alternative wide spread solution to the problem of modifying tuple attributes representing entity properties consists in splitting a tuple into two parts by thus introducing two attribute roles. The first part of a tuple is supposed to be immutable because its main purpose is to identify the entity. The second part of the tuple consists of attributes representing entity properties which are supposed to be mutable. The idea is to somehow reflect the existence of two aspects: (i) entity existence and identification where we only create and delete tuples, and (ii) entity state described by its properties which can change. This approach became integral part of the relational model where "one domain (or combination of domains) of a given relation has values which uniquely identify each element (n-tuple) of that relation and … is called a primary key" [2].

For example, a product item can be represented as a tuple with product id as the only PK attribute while name and price will be non-PK attributes (Fig. 3). If we need to add or remove an entity then we create or delete a whole tuple, respectively. If we need to modify the entity properties, then we update its non-PK attributes (name or price in this example) while PK attributes retain the connection with the entity these properties belong to.

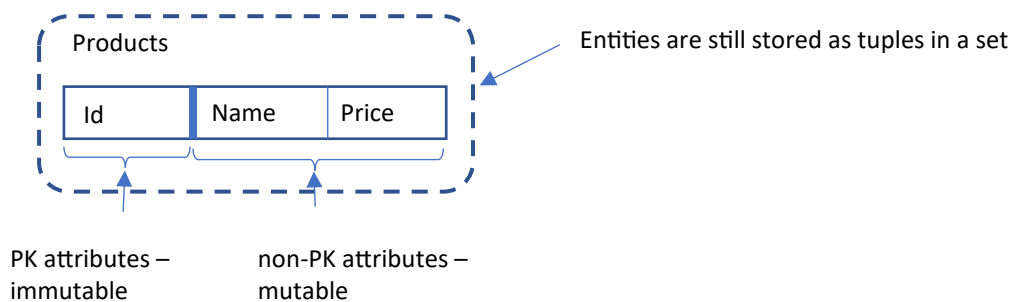

*Fig. 3 A tuple has attributes with two roles: mutable and immutable*

Although this approach is widely used in practice, it has some quite serious problems. Essentially, it retains the original problem but wraps it into a different (conceptual) context and formulations. This original problem – inability to change tuple attributes with the purpose to reflect changing entity properties – now is expressed in different terms.

Probably the best way to formulate this problem is to ask what is a true mathematical tuple in this approach – a whole tuple or only its PK part? Indeed, we introduced two sorts of attributes and hence the question is completely correct. There are at least two possible answers. First, we can assume that a whole tuple with all the original attributes is viewed as a true mathematical tuple, which is a member of the set. Obviously, this assumption does not work because we are not able to modify attributes of true tuples while our non-PK attributes are supposed to be mutable. Second, we can assume that only PK attributes constitute a true mathematical tuple. It is a valid assumption because this part is assumed to be immutable. Yet, now we need to answer the question what are the non-PK attributes? Indeed, if only tuples with PK attributes are members of the set then non-PK part has no (mathematical) status. Non-PK attributes do not have any formal definition and are viewed as some kind of useful

"attachments". Having such a payload in non-PK attributes is very convenient and is actually works very well in practice at least at conceptual level. However, without a clear answer to the question about the formal treatment of non-PK attributes, we lose support of set theory.

Since it is not clear how to interpret the separation on PK and non-PK attributes, we view this mechanism as a workaround or additional conceptual level of description. This trick allows us to switch between practical usefulness and formal correctness by changing our treatment of these parts on the fly depending on what we need. If we need formal mathematical correctness then we say that the whole tuple is a member of the set. If we need practical usefulness then we say that only PK part is a tuple and non-PK part contains mutable properties. All the necessary mathematical formalities are satisfied at the level of full tuples with all the attributes and the PK and non-PK roles is some kind of higher level semantic abstraction. This of course does not solve the problem because it does not change the nature of tuples as immutable values but can be accepted as a justification for many practical use cases and implementations.

A principled solution to this problem which does not conflict with set theory is described in the next section.

## 5 Solution: Representing entity properties by functions

In fact, the semantics of PK attributes correctly reflects one aspect: life cycle of entities and manifestation of the entity existence by supporting only add and remove operations. What is inacceptable in PK-based models is the usage of non-PK attributes: on one hand, they are still tuple attributes and hence are immutable but on the other hand, they are declared mutable. These two requirements are not compatible – we cannot combine mutability with the representation by tuples.

The central idea of our approach is that we completely abandon the usage of tuple attributes, in particular, non-PK attributes, for representing entity (mutable) properties. Instead of artificially attaching non-PK attributes to the true mathematical tuple (or using any other role of attributes), we use mathematical *functions* as an additional construct of the model. A function is interpreted in their original mathematical sense as a *mapping* from one set to another set: $f: X \to Y$ is said to be a function from set $X$ to set $Y$, if and only if $f$ is an operation that assigns to each element $x \in X$, a single element $y = f(x) \in Y$. Functions are then used to represent mutable properties of entities. The main benefit is that functions provide a principled solution to the problem formulated in Section 3 and also significantly simplify the whole model. Now we have only tuples as they are defined in set theory for representing the existence of entities and functions for representing their properties, which can be modified without influencing their existence (and the corresponding tuples).

In our example, product items are represented by tuples with one attribute storing item id (Fig. 4). These tuples are added if a new product item is created and removed if an item is deleted. In terms of RM, relations contain only PKs the attributes of which cannot be modified. In order to represent properties of product items, two functions are created. The first function maps product items to names and the second function maps product items to prices.

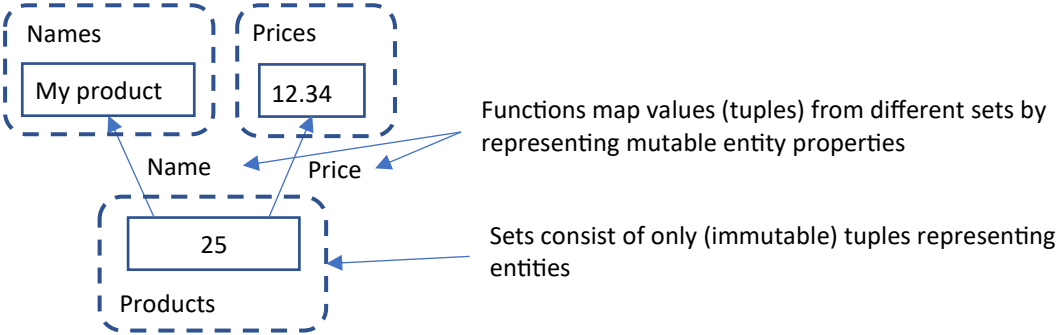

*Fig. 4. Using functions for representing mutable entity properties: instead of modifying tuple attributes, we modify functions*

The main difference of functions in the context of data modeling is that they are supposed to be modified. In contrast to sets which support two basic operations of addition and removal of elements, functions support two other elementary operations: *assigning* a value (setting or writing a value) and *reading* a value (getting the current function output).

Another crucial difference from the purely set-oriented approach is that no tuple attributes (and hence whole tuples) can be modified without any exceptions. If an entity property changes then the corresponding function changes its current output value to some other value. Consequently, the whole database state consists of two parts: (i) the state of sets defined by their elements and (ii) the state of functions defined by their mappings. Here is how the usage of functions changes data modeling:

- A set is defined for each entity type and tuples are used to identify entities by representing the fact of their existence. Sets of the model are modified only if new entities are created or deleted. In terms of the relational model, a set should contain tuples with the structure of traditional PKs excluding all other attributes.
- Entity properties are not represented by tuple attributes – for each entity property a function is defined. If an entity changes its state then some function of this entity is modified. In terms of the relational model, a function should be defined for each FK.

Thus, by means of functions we can solve our problem of modifying immutable tuple attributes. In the next section, we discuss some properties of this model which relies on both sets and functions rather than on only sets.

## 6 Concept-oriented model: both sets and functions for data modeling

The concept-oriented model [5] is a pair $\langle S, F \rangle$ where $S$ is a collection of sets $\{S_1, S_2, \ldots\}$ and $F$ is a collection of functions $\{f_1, f_2, \ldots\}$, $f_i: S_j \to S_k$, $S_j, S_k \in S$. This definition is similar to the *category of sets* from category theory (a category whose objects are sets and morphisms are functions) which could be therefore used as the underlying formalism similar to relational algebra and set theory for the relational model. For comparison, the relational model (its representational part) consists of only sets (relations and domains).

It is a generic definition of COM and depending on additional constraints imposed on the structure of sets and functions we can get more specific models. For example, we could impose constraints on the sets by requiring that they have certain tuple structure (defined by a finite set of attribute types) instead of having the sets storing arbitrary values. We also might work with only one set, the universe of discourse, while functions (all typed by this universal set) will define mappings on this set. In some sense, this model could be thought of as the opposite one to the relational model which works only with sets.

A data *schema* is a model with no set elements and no function elements. In a schema, we specify only sets we are going to use (without members) and function signatures as pairs of their input and output sets. A schema is intended for defining structure all (future) elements must satisfy. Given a schema, the elements of the model define its (current) data state. By adding or removing set elements as well as changing the mappings defined by functions, we can change the current data state of the model. Note that we also can infer new sets and new functions from the model. The main operations for inference are relation composition and function composition.

Below we shortly describe some aspects of this model which help to understand and distinguish it from the conventional set-oriented models relying on only sets.

*Functions store data state.* The database state in COM is determined by both sets and functions rather than only sets. In particular, two databases may have identical sets but different functions. By modifying a mapping between some sets we will get a new state of the database. In this sense, a function is viewed as storage at all levels of organization: conceptual, logical and physical. We can manipulate this (functional) storage by creating and deleting functions as well as modifying the function state, that is, its mapping from inputs to outputs. Note that conventional column stores (column-oriented databases) rely on the physical representation of tables by a number of columns which is very useful for many workloads but does not change the logical and conceptual levels.

*Function is not a binary relation.* A typical argument against independent use and equal role of functions in data modeling is that a function can be formally represented as a binary relation, that is, a set of input-output pairs. In this case, we eliminate the need in functions as a separate notion and hence, presumably, simplify the model because everything including functions can be expressed using sets and set operations. The main problem of this approach is that when representing functions as binary relations we lose the most important aspect (at least for data modeling): the semantics of functions. Indeed, a binary relation by itself does not represent the semantics of mapping one set to another – it is simply a relation. If we still want to have this semantics, that is, the knowledge that the first attribute represents inputs and the second attribute represents outputs, then we again need to introduce additional levels, mechanisms and constraints. Even such simple thing like the constraint that one input may have maximum one output value assigned requires additional mechanisms which make the model more complicated. The situation is actually worse because the whole point of having functions in a data model is that the semantics of mapping is different from the semantics of containment and our goal is to have both of them equally supported. We introduce functions because we postulate that data has two primary aspects: how entities exist and how entities are characterized. The existence is described by sets and tuples. The characterization is described by functions. Removing functions from the model leads to the need in additional mechanisms and level of representation (essentially compensating the absence of functions) like various key types, controversies in typing, difficulties in defining objects, controversies in treating nulls, having no assignment operation and do notion, normalization theory with numerous normal forms etc. Many or maybe all of these problems are consequences of having no functions in the data model.

*Basic semantic relationship.* In any data model, one of the main questions is how data elements are related to each other. In RM, two or more values in some sets are related if there is a tuple which stores these values in its attributes. This relationship is symmetric (all related values have equal status) and has arbitrary arity (we can relate *n* elements). In COM, two values are related if there is a function which maps one of them to the other. This basic relationship is directed and it relates only two values.

*Separating identification and characterization.* Having both sets and functions in a data model reflects two important aspects: 1) existence and identification of entities modeled by tuples and sets with the semantics of containment and add-remove basic operations, and 2) characterization of entities via other entities including primitive values modeled by functions with the semantics of mapping and set-get basic operations. In a purely set-oriented model, it is quite difficult to cleanly separate these two concerns and it is necessary to introduce additional mechanisms and levels which eventually do what functions can do much simpler and more natural.

*Keys and normalization.* Various kinds of keys and numerous normal forms are an integral part of the relational model and probably any set-oriented approach just because using only sets is not enough - something very important will be missing without these mechanisms. On the other hand, ambiguity and controversy of these mechanisms and levels of modeling make the model more complex and difficult to apply. Indeed, there are numerous disputes about the need and usefulness of various kinds of keys and about the usefulness of different normal forms. Introducing functions as one of two primary (semantic, logical and physical) constructs of the model essentially eliminates the need in these mechanisms because now sets have only one purpose: identifying entities and representing what exists. Primary and candidate keys (as an attribute role) are not needed just because any tuple in a set is by definition an identifier of the corresponding entity. Foreign keys are also not needed because functions replace this mechanism by providing many additional benefits like typing, assignment and dot notion. Normalization theory loses its importance as a means of controlling and eliminating redundancy because the problem of redundancy is caused by the need to put data with different roles (identification and characterization) in attributes of one tuple while COM explicitly separates these two roles by putting data in two different constructs: sets and functions. Note that this does not mean that functions somehow magically solve all the problems related to the mechanisms of keys, normalization and functional dependencies (as they are understood in the relational model). Functions allow us to look at these problems from a completely different direction by re-formulating them and finding simpler solutions or even making them obsolete.

*Querying by means of functions.* The introduction of functions as a first-class element of the model is not simply a formal mathematical convenience resolving the issue with immutable tuples. Since functions store data state (along with sets), we can query and process data using functions. In other words, we can infer new data implicitly represented in the database by deriving new functions from existing (maybe also derived) functions. Functions have not only their own representational semantics as mappings – they also have their own operational semantics which relies on the *function composition* operation. Just as the semantics of mapping is opposed to the semantics of containment, the function composition operation is opposed to the relation composition operation (join). The whole general approach to querying and data processing changes: instead of or in addition to inferring new sets in terms of existing sets, we define or infer new functions in terms of existing functions. Some examples where the set-oriented approach is conceptually inadequate and the function-oriented approach solves the problems in a simple and natural way are described in [5], Section 1.1.

*Assignment and dot notion.* The relational model and other set-oriented models have always had difficulties in dealing with assignments just because such an operation does not exist in set theory. Accordingly, dot notion for reading and writing values is also not supported at the same level as set operations. This status contrasts with extremely wide use and semantic clarity of these operations. It is simply difficult to imagine how data can be processed and even thought of without these basic operations. By introducing functions in the model, the status of these operation is accordingly increased and made equal to that of set operations. In general, whenever we think about reading and writing data values, we need to apply functions and operations with functions, and whenever we think about adding and removing data, we need to apply sets and operations with sets.

*Types.* In the relational model, type is defined as "a named, finite set of values" [3]. Yet, "the relational model and type theory are almost completely independent of each other" and type system "complements the relational model" [3]. Essentially, we are able to use the relational model even without strong support of types because types are not inherent part of relational algebra. In particular, relations are not treated as types and cannot be used as types which is strange because relations are sets of values. In contrast, COM treats types as integral part of the model because *any* set is a type and *any* function and attribute must specify its type as some set from this model.

*Objects.* In the relational model, "the very term object itself does not seem to have a precise or universally agreed meaning" [3] which leads to the problem frequently referred to as the object-relational impedance mismatch and the existence of many alternative solutions trying to integrate these concepts. Functions allow us to provide a new definition of an object: an object is a couple of one tuple treated as a reference and a number of function outputs for this reference [5]. Note that an object is not a tuple – only its reference is a tuple. Also, object fields are not a tuple because they exist as values in different sets. In this sense, COM provides another view on this problem especially taking into account ideas (including new mechanism of inheritance) developed in concept-oriented programming [7].

*Other functional models.* Most existing functional models [4, 9, 10] are conceptual models which are closer to the entity-relationship model or they heavily rely on the relational model by translating their constructs and operations to those of relational algebra. In contrast, COM significantly strengthens the role of functions by making them first-class constructs of the model at the same level as sets. In particular, functions are used to represent data state as well as to infer new data. The usage of join is limited by producing a multidimensional space (product) and much less or not at all for relating data values – functions are now used for that purpose. Another important difference is that traditional functional models are typically treated as graph models where sets representing entities are nodes and functions are edges. COM interprets functions (and in general any references) as "member of" or "IS-IN" relation by making it closer to multidimensional models where a set is a multidimensional space with axes defined by its function ranges.

# 7 Conclusion

One of the main postulates behind set-oriented models including the relational model is that data is represented by tuples in sets. This means that all other mechanisms and data interpretations must be expressed in terms of tuples, attributes and sets. In this paper, we described one flaw of this category of models which is a consequence of the immutability of mathematical tuples. The problem is that if tuples are immutable (and they must be immutable if we want to apply set theory), and we assume that entity properties are represented by tuple attributes, then we are not able to modify entity properties in the model without conflict with set theory. In practice, this issue is simply ignored by permitting tuple attribute modifications, that is, if we need to change an attribute then the system just does it. In theory, some theoretical justifications are introduced like remove-add instead of update or assigning special roles to attributes which "legalize" their modification by essentially defining a new layer of the model obeying to different theoretical rules.

In the paper we argued that these workarounds result in a more complicated model with inconsistent theoretical basis and various semantic controversies. The main cause of the problem is the desire to fit both (immutable) entity identifiers and (mutable) entity properties into one construct – tuple. As a principled solution to this problem, we propose to use mathematical functions as first-class elements of the model at the same level as sets. In this model, data is stored not only in sets but also in functions. Moreover, data processing is performed not only by deriving new sets but also by inferring new functions. Such a model based on both sets and functions is much simpler and more natural because it significantly reduces operational and semantic load on sets by producing a nice balance between two aspects: entity existence with add and remove operations modeled by tuples in sets, and entity state with set and get operations modeled by functions. The usefulness of this model was also demonstrated in several open source systems for function-oriented data processing.